\documentclass[10pt,twocolumn,twoside,submit]{JCNtran}

\usepackage[dvips]{epsfig}
\usepackage[latin1]{inputenc}
\usepackage[T1]{fontenc}
\usepackage{amssymb}
\usepackage{amsmath}
\usepackage{bbm}
\usepackage{mathtools}
\usepackage{algorithm}
\usepackage{algorithmic}
\usepackage{tabu}

\DeclarePairedDelimiter\floor{\lfloor}{\rfloor}

\def\BibTeX{{\rm B\kern-.05em{\sc i\kern-.025em b}\kern-.08em
    T\kern-.1667em\lower.7ex\hbox{E}\kern-.125emX}}

\newtheorem{theorem}{Theorem}
\begin{document}
\bibliographystyle{jcn}

\title{Relay Selection in Cooperative Power Line Communication: A Multi-Armed Bandit Approach}
\author{Babak Nikfar and A. J. Han Vinck
\thanks{Manuscript received February 25, 2016; approved for publication by Grace Kim, July 29, 2016.}
\thanks{The authors are with the
Department of Digital Signal Processing, University of Duisburg-Essen, Germany, email: \{babak.nikfar,han.vinck\}@uni-due.de. }} \markboth{JOURNAL OF COMMUNICATIONS AND NETWORKS}{Nikfar \lowercase{\textit{et al}}.: Relay Selection in Cooperative Power Line Communication...} \maketitle

\begin{abstract}
Power line communication (PLC) exploits the existence of installed infrastructure of power delivery system, in order to transmit data over power lines. In PLC networks, different nodes of the network are interconnected via power delivery transmission lines, and the data signal is flowing between them. However, the attenuation and the harsh environment of the power line communication channels, makes it difficult to establish a reliable communication between two nodes of the network which are separated by a long distance. Relaying and cooperative communication has been used to overcome this problem. In this paper a two-hop cooperative PLC has been studied, where the data is communicated between a transmitter and a receiver node, through a single array node which has to be selected from a set of available arrays. The relay selection problem can be solved by having channel state information (CSI) at transmitter and selecting the relay which results in the best performance. However, acquiring the channel state information at transmitter increases the complexity of the communication system and introduces undesired overhead to the system. We propose a class of machine learning schemes, namely multi-armed bandit (MAB), to solve the relay selection problem without the knowledge  of the channel at the transmitter.  Furthermore, we develop a new MAB algorithm which exploits the periodicity of the synchronous impulsive noise of the PLC channel, in order to improve the relay selection algorithm.
\end{abstract}


\section{\uppercase{Introduction}}
\label{1-introduction}

\PARstart{P}{ower} line communication (PLC) is  the technology of transferring data signals through existing power delivery infrastructures, with applications in smart grids, in-vehicle communication, etc. The data signal is generated as a differential voltage between two power delivery conductors, and propagates through the transmission line from source to destination. In power line communication networks, multiple networks {\it nodes} are interconnected via the transmission lines and the data signals flow between different nodes of the network. If the distance between source and destination in a communication scenario is long, the limitation of the transmission range of the PLC node prevents the establishment of a reliable communication. This limitation is due to the harsh environment of the PLC channel, for instance fading, noise, interference, and receiver sensitivity. To overcome this problem, cooperative communication is used to transmit the data signals from source to destination with help of one or more intermediate nodes. In this case, the source node can communicate directly with nodes within its transmission range, and these nodes in turn can forward the message to the destination node. The intermediate nodes are called {\it relays} and the process of transmitting signals with the help of relays is referred to as multi-hop communication or relaying.

The application of relaying in cooperative wireless communication has been studied to a great extent. The use of relaying in cooperative power line communication has been mentioned and  studied for certain communication scenarios as well. For example in \cite{bumiller2002single} and \cite{bumiller2010power}, PLC relaying based on single frequency networking has been introduced and its performance has been discussed. Distributed space-time coding for multi-hop transmission and decode-and-forward relaying has been studied in \cite{lampe2006distributed} and \cite{tonello2010opportunistic}, respectively. The concept of cooperative multi-hop communication for PLC has been first introduced and discussed in \cite{balakirsky2005potential}, \cite{lampe2011cooperative} and \cite{lampe2012cooperative}. Existence of many intermediate nodes between source and destination, results in the existence of many optional paths or routes to follow. This situation gives rise to the problem of proper relay selection, where the challenge is to pick the optimal path that satisfies the needed performance requirements. In \cite{lampe2012cooperative} the relay selection problem has been discussed and the link rate has been introduced as a figure of merit for different relay selection criteria.

In this paper, we consider the two-hop cooperative communication scenario, in which the transmitted signal from source travels to an intermediate relay before reaching the destination. We consider $N$ available intermediate relay nodes, from which one of them as the relaying node is to be selected. However, the proper relay selection policy requires the availability of channel side information at the transmitter, which in turn requires an increased complexity of signal processing and introduces a lot of overhead in the system. In order to avoid this problem, we introduce a class of machine learning algorithms, namely {\it multi-armed bandit (MAB)}, which helps us find the best relay from available relays, based on the performance of the channel in both of the communications hops. Different algorithms of MAB has been developed in the field of machine learning, for example \cite{agrawal1995sample,kocsis2006discounted,garivier2008upper}. In wireless communication, MAB has been used in order to solve problems which are dealing with acquiring the best selection policy. For example, in \cite{maghsudi2013relay} and \cite{maghsudi2013relay2}, MAB has been used for relay selection problem in a stationary wireless channel. In PLC, MAB has been introduced in \cite{mypaper} to solve the channel selection problem in a multichannel PLC system. In this paper, we introduce the MAB tool to solve the relay selection problem in two-hop cooperative PLC without access to the channel side information. Furthermore, we propose two new algorithms based on MAB to further improve the performance of the relay selection based on the channel characteristics of the power line systems. The proposed algorithms are adapted in such way to exploit the specific features of the PLC channel, namely the periodicity of the synchronous impulsive noise of the network. This adaptation to PLC of the proposed algorithms proves to be advantageous as illustrated in the numerical results.

This paper is organized as follows. Section \ref{2-chapter} describes the system and channel model of the PLC network. The cooperative PLC and the relay selection problem are presented in Section \ref{3-chapter}. Multi-armed bandit model is introduced in Section \ref{4-chapter} as well as the representation of the existing MAB algorithms. The proposed algorithms are described in Section \ref{5-chapter} and simulation results are presented in Section \ref{6-chapter}. Finally, Section \ref{7-conclusion} concludes the paper.

\vspace{10pt}
\section{\uppercase{System Model}}
\label{2-chapter}

\begin{figure}[t]
\begin{center}
\epsfxsize=6cm \leavevmode\epsfbox{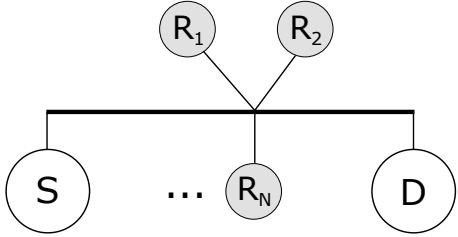} \caption{PLC network with a transmitter, a receiver, and $N$ relay nodes.} \label{Fig:plcNetwork}
\end{center}
\end{figure}

Power line communication channels can be characterized as frequency-selective time-variant channels. We consider a PLC network, consisting of a transmitter node, a receiver node, and $N$ intermediate nodes known as relays. In PLC networks, due to line attenuations, fading, noise, and signal interferences, transmission over long distances results in a degraded performance which may lie below the performance requirements of the system. Therefore, each node can communicate directly with those nodes which fall into its transmission range. These intermediate nodes transmit the message to the destination, constructing a two-hop communication. The structure of a two-hop communication can be seen in Figure \ref{Fig:plcNetwork}, where S and D represent source (transmitter) and destination (receiver), respectively, and $R_i,~i\in\{1,\cdots,N\}$ are $N$ available relay nodes which can help the transmission of PLC signals from transmitter to receiver.

We use the bottom-up deterministic approach in PLC channel modeling, namely the transmission line theory model. Therefore, we can model each two nodes of the PLC network with voltages $V_i(f)$ and $V_j(f)$, respectively, and corresponding currents $I_i(f)$ and $I_j(f)$, with ABCD-parameters as follows \cite{lampe2011cooperative,lampe2012cooperative}. Note that both voltages and currents are frequency dependent.
\begin{equation}
	\left[ \begin{matrix} V_i(f) \cr I_i(f) \end{matrix} \right] = \left[ \begin{matrix} A(f) & B(f) \cr C(f) & D(f) \end{matrix} \right] \left[ \begin{matrix} V_j(f) \cr I_j(f) \end{matrix}  \right].
\end{equation}
The corresponding transfer function between these two nodes can be derived as
\begin{equation}
	H_{ij}(f) \triangleq \frac{V_j(f)}{V_i(f)} = \frac{Z_j(f)}{A(f)Z_j(f)+B(f)},
\end{equation}
where $Z_j(f) = V_j(f)/I_j(f)$ is the impedance of node $j$. The transfer function of the channel between two network nodes $i$ and $j$, with an intermediate node $k$ in between, can therefore be calculated as
\begin{equation}
	H_{ij}(f) = H_{ik}(f)\cdot H_{kj}(f).
\end{equation}

ABCD-parameters or transmission line parameters depend on the power line characteristics as well as the length $\ell$ of the transmission line, and can be calculated as
\begin{equation}
	\left[ \begin{matrix} A(f) & B(f) \cr C(f) & D(f) \end{matrix} \right] = 
	\left[ \begin{matrix} \cosh(\gamma\ell) & Z_0\sinh(\gamma\ell) \cr \frac{1}{Z_0}\sinh(\gamma\ell) & \cosh(\gamma\ell) \end{matrix}  \right],
\end{equation}
where $Z_0 = \sqrt{(R+j2\pi fL)/(G+j2\pi fC)}$ is the characteristic impedance of the transmission line per unit length and $\gamma = \sqrt{(R+j2\pi fL)(G+j2\pi fC)}$ is the propagation constant, and they are related to the primary cable parameters R, G, L, and C, representing resistance, conductance, inductance, and capacitance of the line per unit length. The primary cable parameters depend on the physical cable characteristics, and are derived for a PLC channel in \cite{7147610}. The propagation constant is a complex quantity and can be written as $\gamma = \alpha(f) + j\beta(f)$, where $\alpha(f)$ is the attenuation constant, and $\beta(f)$ is the phase constant. The channels in this model are considered to be independent and channel correlation is neglected. This assumption is due to independent loads in a PLC network and the geographically distributed routs of a relay network which makes the independent channel assumption reasonable. 

In in-home PLC applications, dominant noise components consist of background noise, aperiodic impulsive noise, as well as synchronous and asynchronous impulsive noise. A prominent noise in PLC is the periodic impulsive noise which is synchronous to the  AC (alternating current) of the mains \cite{ferreira2011power}. Therefore, the noise is considered to be {\it cyclostationary} and the periodic instantaneous noise power is derived in \cite{katayama2006mathematical} as
\begin{equation} \label{eq:noisevar}
	\sigma^2_N(t) = \sum_{l=0}^{L-1} A_l \left| \sin\left( \frac{2\pi t}{T_{AC}} + \theta_l \right)\right|^{n_l},
\end{equation}
where $L$ represents the number of noise classes (for narrowband PLC, $L=3$ \cite{katayama2006mathematical}), $A_l$, $\theta_l$, and $n_l$ are different characteristic parameters of the $l$-th noise class, and $T_{AC}$ is the period of the mains voltage. 

\vspace{10pt}
\section{\uppercase{Cooperative PLC and Problem Formulation}}
\label{3-chapter}

\begin{figure}[t]
\begin{center}
\epsfxsize=8cm \leavevmode\epsfbox{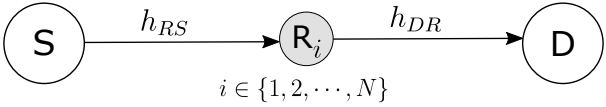} \caption{A two-hop transmission in cooperative PLC.} \label{Fig:twoHop}
\end{center}
\end{figure}

The idea of cooperative communication has been well investigated in wireless communication, e.g. in \cite{nosratinia2004cooperative,laneman2004cooperative}. The principle of this idea is to realize spatial diversity without the use of multiple antennas. In this case, cooperative users generate a virtual antenna array to achieve the desired cooperative diversity. This concept has been extended to relay networks with multi-hop  transmission between source and destination, e.g. in \cite{boyer2004multihop}. The concept of cooperative diversity has been introduced in PLC for the first time in \cite{balakirsky2005potential}, and has been further discussed in \cite{lampe2011cooperative,lampe2012cooperative}. It has been shown in \cite{lampe2012cooperative} that the PLC relay channel consists of two keyhole channels, and thus a diversity gain as observed for wireless relaying cannot be achieved for PLC. However, despite the lack of the cooperative diversity advantage, cooperative multi-hop transmission can provide significant power gains. In this paper, we assume a two-hop transmission, that is, a source node transmits the message to a destination node through a relay node between them as depicted in Figure \ref{Fig:twoHop}. A generalization to a multi-hop transmission is straightforward.

In a two-hop cooperative PLC, as depicted in Figure \ref{Fig:twoHop}, three nodes are available, namely source node $S$, selected relay node $R_i$, which is selected from a sequence of all the available relays $i\in\{1,2,\cdots,N\}$, and destination node $D$. We assign a number $n$ to each node so that $\{S,R_i,D\}\thicksim\{1,2,3\}$. The cooperative transmission is assumed to follow a time division protocol, which means at each time instant, only one node (either the source or the relay) can transmit data with a fixed transmit power. As a figure of merit, we consider the end-to-end achievable rate, also known as the {\it end-to-end capacity}. The end-to-end capacity in a two-hop transmission for a link between node $n$ and node $n+1$ is expressed as $C_{n,n+1}$,  which is a strictly increasing function of the transmitted signal-to-noise ratio (SNR) \cite{ferreira2011power}. The SNR performance of the transmission links directly exhibits the overall performance of transmission, hence forming a reliable figure of merit.

We use the conventional strategy of {\it fixed-rate two-hop} transmission \cite{oyman2007multihop}. The reason to choose the fixed-rate strategy is due to the fact that in other two-hop strategies such as rate-adaptive scheme, the link capacities are required before transmission so that the transmitter can adjust the rate to follow the link capacity. In a fixed-rate cooperative transmission scheme, the relay node between the source and destination nodes, retransmits the received message from source using the same transmission scheme and thus link rate.  Hence, the maximum rate that can be achieved over this route is determined by the minimum of the rates achievable on the individual links. Let us denote the data rate over each hop as $R_1=R_2=R$, for some fixed value of $R$. In order to ensure reliable communication, $R\leq C_{n,n+1}$ must be satisfied over all hops. It is shown that maximum end-to-end capacity can be achieved by choosing $R=\min_{n=1,2}C_{n,n+1}$ \cite{oyman2007multihop}. The end-to-end capacity for a fixed-rate two-hop transmission can be expressed as
\begin{equation}
	C_{total} = \frac{1}{2} \min_{n=1,2} C_{n,n+1}.
\end{equation}

Due to the time-variant nature of the PLC systems, in order to follow the changes of the channel and selecting the best relay, channel state information (CSI) should be available at the transmitter at all times. However, acquiring CSI requires constant  transmission of pilot signals throughout the transmission which in turn enormously increases the overhead and decreases the throughput. Moreover, channels formed by all of the available relays must be evaluated in order to perform an optimal relay selection, which with a high number of available relays is not an easy task. We wish to have a relay selection strategy which results in a high end-to-end capacity $C_{total}$. Formally, we aim to solve the following maximization problem
\begin{equation}
	\underset{R_i\in\{R_1,R_2,\cdots,R_N\}}{\textup{maximize}}~~ C_{total}.
\end{equation}

We consider three relay selection strategies to maximize the total end-to-end capacity without any information at transmitter.
\begin{enumerate}
	\item Fixed selection: in this strategy, a fixed relay node is assigned for cooperative transmission between the source and destination, regardless of the instantaneous channel conditions. This strategy neglects the variations is the PLC channel and therefore is not an optimal method of relay selection.
	\item Random selection: in this strategy, at each transmission time interval a random relay is selected from the sequence of $N$ arrays. This method neglects the variations of the channel over time as well; however, the randomness of the relay selection may decrease the probability of selecting a bad relay compared to the fixed selection method.
	\item The proposed learning algorithms: we propose two new learning algorithms based on the {\it multi-armed bandit (MAB)} model in machine learning. In our approach we consider the variations of the channel over time which occurs in a cyclostationary manner as described  before, and try to adapt the relay selection with these variations without the knowledge of CSI at the transmitter. The concept of multi-armed bandit and our algorithmic solutions are discussed in the following sections. Furthermore, we show through numerical results that our proposed algorithms result in a better performance compared to other relay selection approaches.
\end{enumerate} 

\vspace{10pt}
\section{\uppercase{Multi-armed Bandit Problem Modeling}}
\label{4-chapter}

Multi-armed bandit (MAB) is a class of decision making problems introduced in machine learning field, where an agent sequentially selects an arm (action, interchangeably) from a set of predefined arms (actions), and receives a reward drawn from some a priori unknown distribution. Only the reward of played arm is observable. As a result of lack of information, at each trial, the player may choose an inferior arm in terms of average reward, yielding a {\it regret} that is quantified by the difference between the reward that would have been achieved if the agent would have selected the best arm with the highest reward and the actual achieved reward.

We model the PLC relay selection problem as a MAB. Let us assume a set $\mathcal{K}$ of available relay nodes, here and thereafter known as {\it actions} or {\it arms}. Each frame of data is transmitted by selecting a relay node $k\in\mathcal{K}$, $k=\{1,2,\cdots,N\}$, resulting in a particular total end-to-end capacity, here and thereafter known as {\it reward}. At each time slot $t$ (corresponding to one frame of data), an action $a_t$ is selected, yielding the instantaneous reward $X_t(a_t)$. The rewards $\{X_t(k)\}_{t\geq 1}$ for each arm $k\in\{1,2,\cdots,N\}$ is calculated according to the received signal and this information is fed back to the transmitter via a robust mode of transmission and the transmitter chooses an arm at each trial according to a policy $\pi$. Let us denote the expectation of the reward $X_t(k)$ by $\mu_{k,t}$. Let $k^*_t$ denote the optimal arm at time t, with expected reward $\mu_{k^*,t}$, where by definition $\mu_{k^*,t} = \mu^*_t = \underset{k\in\mathcal{K}}{\textup{max}}~~ \mu_{k,t}$. We define the {\it instantaneous regret} at time $t$ as the difference between mean rewards of the selected arm and the optimal arm. The expected regret of a decision making policy $\pi$ after $T$ trials, therefore, can be expressed as
\begin{equation}
	R_{\pi,T}={E}_{\pi} \left[\sum_{t=1}^T \left(\mu^{*}_{t}-\mu_{a_{t},t} \right) \right],
\end{equation}
where ${E}[.]$ represents the mathematical expectation. The goal of a good policy is to select the optimal arm at each trial, which results in a minimum expected regret over all trials. Therefore, the goal of the MAB problem is to minimize the expected regret with a certain decision making policy $\pi$, or equivalently
\begin{equation}
	\underset{\pi}{\textup{minimize}}~R_{\pi,T}.
\end{equation}

\subsection{UCB and modified UCB algorithms}

{\it Upper-Confidence Bound} (UCB) algorithms are deterministic MAB policies which have been introduced and analyzed by \cite{agrawal1995sample}. In the seminal UCB algorithm, an upper-bound for the expected reward $\mu_{k,t}$ is evaluated which its calculation is based on the previous rewards of that particular arm and some uncertainty factor. During the $t$-th round, the user selects an arm which maximizes the upper-bound of the confidence interval for expected reward $\mu_{k,t}$. The algorithm starts the selection process by the initialization phase, in which during the first $N$ rounds, action $a_k$, $k\in\{1,2,\cdots,N\}$ is selected successively and after transmission at each arm, the corresponding UCB index $I_{k,t}$ of that arm is calculated as
\begin{equation} \label{eq:index}
	I_{k,t} = \bar{X}_t(k) + c_t(k),
\end{equation}
where
\begin{equation}
	\bar{X}_t(k) = \frac{1}{N_t(k)} \sum_{s=1}^t X_s(k) \mathbbm{1}_{\{a_s=k\}}
\end{equation}
is the empirical mean of the previous rewards up to the time $t$, and $N_t(k)=\sum_{s=1}^t \mathbbm{1}_{\{a_s=k\}}$ is the number of times arm $k$ has been selected up to time $t$, and $\mathbbm{1}_{\{a_s=k\}}$ returns one if $a_s=k$ and zero otherwise. The second term in (\ref{eq:index}) is referred to as a padding function. It describes the uncertainty factor of the corresponding arm, which has high values for less selected arms and vice versa. The purpose of the padding function is to ensure an exploration-exploitation policy in which the less we have played an arm, the more uncertain we are about the calculated empirical mean and thus, a bigger padding function. This results in selecting the less selected arms due to their bigger padding functions, hence explore the other arms. A standard choice for the padding function is 
\begin{equation}
	c_t(k) = B \sqrt{\frac{\xi\log(t)}{N_t(k)}},
\end{equation}
where $B$ is an upper-bound on the rewards and is selected as the maximum value of the observed reward through many trials of the algorithm. $\xi>0$ is a parameter which is used to tune the algorithm to obtain the best results and is selected based on empirical applications of the algorithm \cite{garivier2008upper}. Finally, for the next rounds, the arm yielding the maximum UCB index is selected and after the transmission through the selected arm, the corresponding index is updated. The selection policy for UCB, therefore, can be expressed as
\begin{equation}
	a_t = \underset{1\leq k\leq N}{\textup{arg~max}}~\bar{X}_t(k) + c_t(k).
\end{equation}

\begin{theorem}
\label{th:ucb}
Upper-confidence bound algorithm is optimal in the sense that its expected regret matches the lower bound regret of all policies for stationary bandit problems \cite{agrawal1995sample}.
\end{theorem}

Theorem \ref{th:ucb} denotes the optimality of UCB policy for stationary bandits. However, for non-stationary bandits UCB cannot be considered as an optimal policy. For non-stationary bandits the rewards are assumed to be non-stationary, therefore the optimal policy must have the ability of adaptation to the changes of the statistical characteristics of the rewards. For piece-wise stationary bandits, {\it discounted upper-confidence bound} (D-UCB) algorithm has been introduced in \cite{kocsis2006discounted}, where a discount factor $\gamma\in(0,1)$ has been introduced to the UCB algorithm to mark the effects of the time in which the selected arm has been played. This means the more past actions do not have equal weights in calculating the empirical mean of the rewards, whereas the more recent selected actions weigh more in the calculation of the confidence bound index. Details of the D-UCB algorithm is described in Algorithm \ref{alg:DUCB}.

\begin{algorithm}
\caption{Discounted Upper-Confidence Bound}
\label{alg:DUCB}
\small
	\begin{algorithmic}[1]
	\FOR {$t=1,...,N$}
	 	\STATE Let $a_{t}=t$, play and observe the reward.
		\STATE Calculate the discounted empirical mean as 
	 			$$ \bar{X}_t(\gamma,k) = \frac{1}{N_t(\gamma,k)} \sum_{s=1}^t \gamma^{t-s} X_s(k) \mathbbm{1}_{\{a_s=k\}},$$ $$N_t(\gamma,k) = \sum_{s=1}^t \gamma^{t-s} \mathbbm{1}_{\{a_s=k\}} $$
		\STATE Calculated the discounted padding function as
				$$ c_t(\gamma,k) = 2B \sqrt{\frac{\xi\log(n_t(\gamma))}{N_t(\gamma,k)}},$$ $$n_t(\gamma) = \sum_{k=1}^N N_t(\gamma,k) $$
		\STATE Update the discounted UCB index $I_{k,t,\gamma} = \bar{X}_t(\gamma,k) + c_t(\gamma,k)$ of the arm.
	\ENDFOR
	\FOR {$t=K+1,...,T$}
		\STATE Let $a_{t}= \underset{1\leq k\leq N}{\textup{arg max}}~\bar{X}_t(\gamma,k) + c_t(\gamma,k)$, play and observe the reward.
		\STATE Update the discounted UCB index of the selected arm.
	\ENDFOR
	\end{algorithmic}
\end{algorithm}

\begin{theorem}
\label{th:ducb}
Discounted upper-confidence bound algorithm is almost optimal in the sense that its expected regret matches the lower bound regret of all policies for piece-wise stationary bandit problems \cite{kocsis2006discounted,garivier2008upper}.
\end{theorem}

Theorem \ref{th:ducb} denotes that the D-UCB algorithm is proved to be almost optimal for piece-wise stationary bandits. However, in PLC network relay selection problem, the reward of each arm is denoted by the end-to-end capacity of that arm, which in turn depends inversely on the noise power. As mentioned in Section \ref{2-chapter}, the noise in PLC networks is a cyclostationary process. We propose in this paper, two more appropriate variants of UCB algorithm, designed specifically for cyclostationary behavior of the PLC channel.

\vspace{10pt}
\section{\uppercase{The Proposed Algorithmic Solutions}}
\label{5-chapter}

In a cyclostationary process, the statistical characteristics of the process repeat periodically. Therefore, in the calculation of the empirical mean and the padding function, considering all the previous actions with the same weight, as in UCB, is not an optimal policy. Furthermore, the mere consideration of the recent past actions as major contributors to the calculation of the confidence index, as in D-UCB, may result in a sub-optimal policy as well, since the far past actions in the same cycle and hence with the same statistical characteristics are neglected due to their low discount weight. To overcome this problem, we propose two novel algorithms, namely {\it cyclo-discounted upper-confidence bound} algorithm and {\it sinusoidal upper-confidence bound} algorithm. Furthermore, through simulation results, we show that for a cyclostationary system like PLC, these algorithms result in a better selection policy, and therefore a better performance.

\subsection{The Proposed Cyclo-Discounted UCB Algorithm}

\begin{figure}[t]
\begin{center}
\epsfxsize=9cm \leavevmode\epsfbox{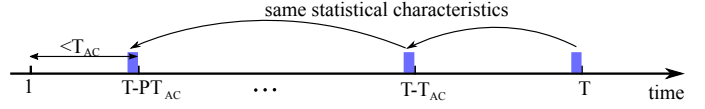} \caption{The cyclostationary behavior of the channels with mains period $T_{AC}$.} \label{Fig:proposed}
\end{center}
\end{figure}

Let us assume the period of the AC waveform of the power lines as $T_{AC}$ with the noise power and hence the rewards of each arm as a cyclostationary process with $T_{AC}$ duration of each cycle. Up to the time index $t=T$, the total number of complete cycles, denoted by $P$, can be calculated as
\begin{equation} \label{eq:P}
	P = \floor*{\frac{T}{T_{AC}}}.
\end{equation}
Furthermore, the empirical mean value of each arm at time $t$, as well as the padding function, are to be calculated in a way that all the last cycles are included in the calculation. The weighing factor should be chosen in a way that it involves the cyclostationary behavior of the reward. For this purpose, let us consider a single period of time with duration $T_{AC}$. In this period, the first samples will have lower weights and the last samples will have higher weights. In other words, we use a D-UCB discount factor for each period. For the last $P$ complete periods, we calculate the empirical means as
\begin{eqnarray} \label{eq:Xi2}
	\Xi_2 &=& \Xi_2(t,T_{Ac},\gamma,k)  \\
	 &=& \sum_{p=1}^P \sum_{s=t-pT_{AC}}^{t-(p-1)T_{AC}} \gamma^{t-(p-1)T_{AC}-s} X_s(k) \mathbbm{1}_{\{a_s=k\}}, \nonumber
\end{eqnarray}
where the term $\gamma^{t-(p-1)T_{AC}-s}$ applies a discounted factor for each period separately. For the incomplete period at the beginning of the time index (as depicted in Figure \ref{Fig:proposed}), we calculate the corresponding portion of the empirical mean as
\begin{eqnarray} \label{eq:Xi1}
	\Xi_1 &=& \Xi_1(t,T_{Ac},\gamma,k) \\
	 &=& \sum_{s=1}^{t-PT_{AC}} \gamma^{t-PT_{AC}-s} X_s(k) \mathbbm{1}_{\{a_s=k\}}. \nonumber
\end{eqnarray}
Therefore, for $1\leq t\leq T$,  the empirical mean can be calculated as
\begin{equation} \label{eq:Xi1Xi2}
	\bar{X}_t(T_{AC},\gamma,k) = 
	\frac{\Xi_1+ \varkappa ~ \Xi_2}{N_t(T_{AC},\gamma,k)}
\end{equation}
where $\varkappa=sign(P)$, and $N_t(T_{AC},\gamma,k)$ is described as
\begin{eqnarray} \label{eq:Xi1Xi2N}
	N_t(T_{AC},\gamma,k) &=& 
	\sum_{s=1}^{t-PT_{AC}} \gamma^{t-PT_{AC}-s}  \mathbbm{1}_{\{a_s=k\}} \\ && + \varkappa 
	\sum_{p=1}^P \sum_{s=t-pT_{AC}}^{t-(p-1)T_{AC}} \gamma^{t-(p-1)T_{AC}-s} \mathbbm{1}_{\{a_s=k\}}. \nonumber
\end{eqnarray}

The equations (\ref{eq:Xi2}), (\ref{eq:Xi1}), (\ref{eq:Xi1Xi2}), and (\ref{eq:Xi1Xi2N}) are used to calculate the empirical mean so that the periodic elements at each cycle have more contribution to the final value. This results in a more accurate empirical mean for a cyclostationary process. Furthermore, the padding function is calculated in the same way as before, but with the new value of $N_t(T_{AC},\gamma,k)$ which contains the cyclostationary weighing method as well. Finally, at time $T$, the arm with the highest UCB index is selected as the next route for transmission. This algorithm is summarized in Algorithm \ref{alg:CDUCB}.

\begin{algorithm}
\caption{Proposed Cyclo-Discounted Upper-Confidence Bound}
\label{alg:CDUCB}
\small
	\begin{algorithmic}[1]
	\FOR {$t=1,...,N$}
	 	\STATE Let $a_{t}=t$, play and observe the reward.
		\STATE Calculate the discounted empirical mean $\bar{X}_t(T_{AC},\gamma,k)$ as described in (\ref{eq:Xi2}), (\ref{eq:Xi1}), (\ref{eq:Xi1Xi2}), and (\ref{eq:Xi1Xi2N}).
		\STATE Calculated the discounted padding function as
				$$ c_t(T_{AC},\gamma,k) = 2B \sqrt{\frac{\xi\log(n_t(T_{AC},\gamma))}{N_t(T_{AC},\gamma,k)}},$$ $$n_t(T_{AC},\gamma) = \sum_{k=1}^N N_t(T_{AC},\gamma,k) $$
		\STATE Update the cyclo-discounted UCB index $I_{k,t,\gamma,T_{AC}} = \bar{X}_t(T_{AC},\gamma,k) + c_t(T_{AC},\gamma,k)$ of the arm.
	\ENDFOR
	\FOR {$t=K+1,...,T$}
		\STATE Let $a_{t}= \underset{1\leq k\leq N}{\textup{arg max}}~\bar{X}_t(T_{AC},\gamma,k) + c_t(T_{AC},\gamma,k)$, play and observe the reward.
		\STATE Update the cyclo-discounted UCB index of the selected arm.
	\ENDFOR
	\end{algorithmic}
\end{algorithm}

\begin{figure*}[t]
\begin{center}
\epsfxsize=16cm \leavevmode\epsfbox{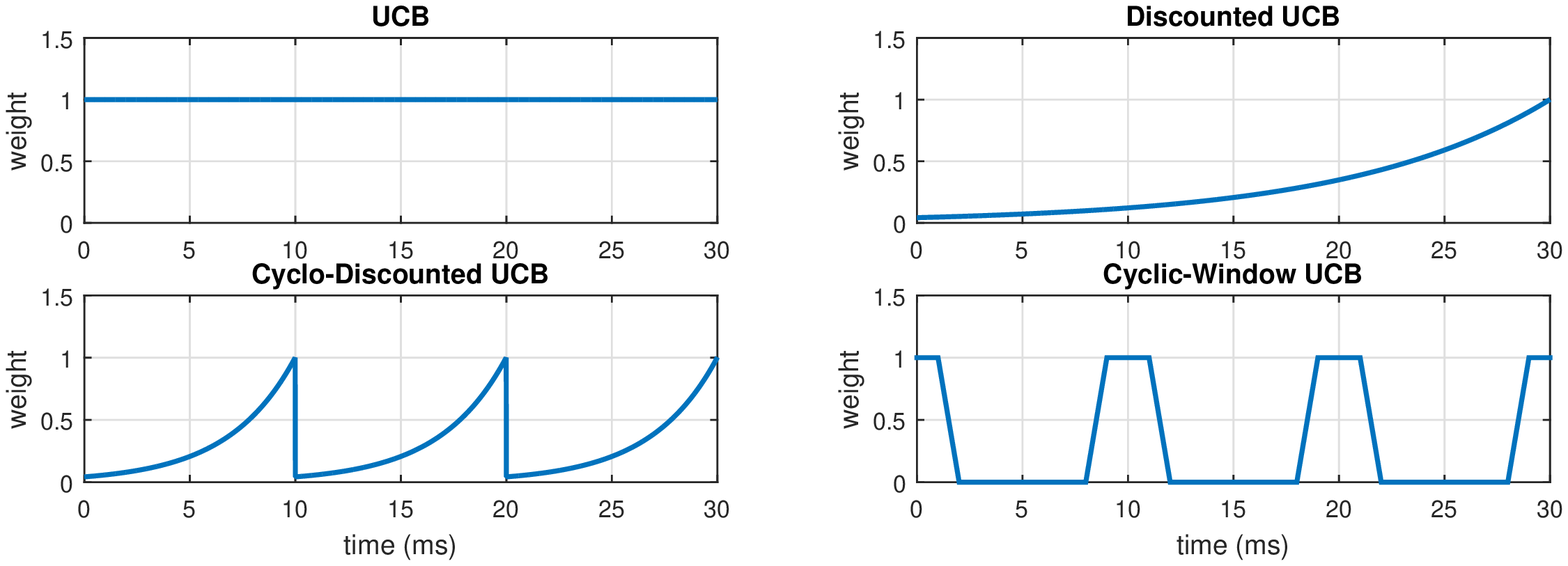} \caption{Weight factors in confidence index calculation for UCB, discounted UCB, proposed cyclo-discounted UCB, and proposed cyclic-window UCB algorithms.} \label{Fig:weight}
\end{center}
\end{figure*}

\subsection{The Proposed Cyclic-Window UCB Algorithm}

Similar to the approach in cyclo-discounted UCB algorithm, we propose another algorithm which its weight factor is periodic, hence adapted to the cyclostationary behavior of the channel. In cyclo-discounted UCB algorithm, the statistical characteristics of the reward function at time $t=T$ and $t=T-T_{AC}$ are the same, and therefore the weight factor at these times is at its maximum. However, at time $t=T-T_{AC}+1$, the reward function enters the next cycle and results in a low value of weight factor. Although, the statistical characteristics of the reward function at time $t=T-T_{AC}$ and $t=T-T_{AC}+1$ are not that different. This problem can be addressed with the proposed {\it cyclic-window UCB} algorithm. 

In cyclic-window UCB algorithm, the empirical mean and the padding function are not discounted as in cyclo-discounted UCB, but are windowed periodically. The windowing period is chosen to be matched with the periodic behavior of the reward and hence matched with $T_{AC}$. The window size, $W$, is selected to maximize the effect of windowing. Formally, we can calculate the empirical mean as
\begin{eqnarray} \label{eq:sucb1}
	\bar{X}_t(W,T_{AC},k) &=& \frac{1}{N_t(W,T_{AC},k)}  \sum_{p=0}^{P} \sum_{s=1}^t  w(s-t+pT_{AC})  \nonumber \\ && X_s(k) \mathbbm{1}_{\{a_s=k\}},
\end{eqnarray}
where $P$ is defined in (\ref{eq:P}), and $w(s)$ is the window function and is defined as
\begin{equation} \label{eq:window}
	w(s) = \left\{\begin{array}{ll}1&|s|<\frac{W}{2}\\0&\mathrm{otherwise}\end{array}\right.
\end{equation}
The term $w(s-t+pT_{AC})$ in (\ref{eq:sucb1}) denotes that the windowing is performed at the current time in addition to the multiple times of the mains period before the current time (see Figure \ref{Fig:weight}). The term $N_t(W,T_{AC},k)$ can be described as
\begin{equation} \label{eq:sucb2}
	N_t(W,T_{AC},k) = \sum_{p=0}^{P}  \sum_{s=1}^t w(s-t+pT_{AC})  \mathbbm{1}_{\{a_s=k\}}.
\end{equation}
The weight factors in confidence index calculation of the proposed algorithms as well as that of the basic UCB and discounted UCB algorithms are depicted in Figure \ref{Fig:weight}. The proposed cyclic-window UCB  algorithm is summarized in Algorithm \ref{alg:SUCB}.

\begin{algorithm}
\caption{Proposed Cyclic-Window Upper-Confidence Bound}
\label{alg:SUCB}
\small
	\begin{algorithmic}[1]
	\FOR {$t=1,...,N$}
	 	\STATE Let $a_{t}=t$, play and observe the reward.
		\STATE Calculate the discounted empirical mean $\bar{X}_t(W,T_{AC},k)$ as described in (\ref{eq:sucb1}), (\ref{eq:window}), and (\ref{eq:sucb2}).
		\STATE Calculated the discounted padding function as
				$$ c_t(W,T_{AC},k) = 2B \sqrt{\frac{\xi\log(n_t(W,T_{AC}))}{N_t(W,T_{AC},k)}},$$ $$n_t(W,T_{AC}) = \sum_{k=1}^N N_t(W,T_{AC},k) $$
		\STATE Update the cyclic-window UCB index $I_{k,t,W,T_{AC}} = \bar{X}_t(W,T_{AC},k) + c_t(W,T_{AC},k)$ of the arm.
	\ENDFOR
	\FOR {$t=K+1,...,T$}
		\STATE Let $a_{t}= \underset{1\leq k\leq N}{\textup{arg max}}~\bar{X}_t(W,T_{AC},k) + c_t(W,T_{AC},k)$, play and observe the reward.
		\STATE Update the cyclic-window UCB index of the selected arm.
	\ENDFOR
	\end{algorithmic}
\end{algorithm}

The proposed algorithms consist of an initialization phase which its duration is proportional to the number of relays. Then at each frame length the end-to-end capacity has to be obtained and instantaneous rewards are calculated at the transmitter. The calculation of reward for the selected arm consists of a linear calculation of the empirical mean in addition to the calculation of the padding function which consists of a square root and a logarithm function. These calculations are only done in one arm at each frame time. The algorithms are bounded in the sense that after a limited amount of time, the arm with the best rewards can be detected and selected for most of the consecutive selections. However, the amount of time needed for this convergence, denoted by $\tau_{alg}(N)$, is directly proportional to the number of relays $N$ (see Figure \ref{Fig:sim6}). Let us denote the time in which the relay channels remain in a particular state as $\tau_{ch}$. In order to have a working algorithm, change of the channel in time must happen slower than the convergence time of the algorithm and $\tau_{alg}(N)<\tau_{ch}$ must hold. The exhaustive search method in the simulations, assumes the perfect CSI at transmitter and expectedly the returned reward is higher than learning algorithms. However, acquiring CSI is much more complicated than feeding back the observed reward. The reason for that is that the PLC channel is time-variant and frequency-selective. Therefore, pilot signals should be transmitted in all the subcarriers at pre-defined time intervals throughout the transmission. Moreover, the amount of overhead which this brings increases linearly by the number of available relays, since in order to react to the changes in the environment all the possible routs should be evaluated. On the other hand, feeding back the reward data can take place on the free bits of the ACK (acknowledgment) packet which is already being fed back to the transmitter and is not dependent on the number of available relays.

\vspace{10pt}
\section{\uppercase{Numerical Results}}
\label{6-chapter}

We consider a two-hop cooperative communication with a PLC channel as described in Section \ref{2-chapter}. The OFDM parameters of the PLC system is listed in Table \ref{tab:parameters}. The number of available relays is considered to be 6 nodes, from which a single node is selected for transmission at each transmission instant. The only difference between the available relay nodes is the corresponding link rate, which in turn is dependent on the channel response as well as the noise power spectral density of the PLC channel. Given the frequency response $H_{n,n+1}(f)$ for a link from node $n$ to node $n+1$, the link rates are computed as \cite{lampe2012cooperative}
\begin{equation} \label{eq:linkr}
	C_{n,n+1} = \int_{f_1}^{f_2} \log_2 \left(1+\frac{S_T|H_{n,n+1}(f)|^2}{N_0\Gamma}\right) \mathrm{d}f,
\end{equation}
where $S_T$ is the transmitter-side power spectral density, $N_0$ is the receiver-side noise power spectral density, and $\Gamma=10$ is the margin taking into account the gap between information-theoretic capacity and achievable rate using practical coding and modulation schemes.

\begin{table}[!h]
\begin{center}
\begin{tabu} to 0.5\textwidth {|X|c|}
	\hline
	\textbf{Parameters} & \textbf{Value} \\
	\hline
	Number of subcarriers (samples) & 128 \\
	\hline
	Number of used subcarriers (samples) & 102 \\
	\hline
	Number of cyclic prefix (samples) & 30 \\
	\hline
	OFDM interval ($\mu s$) & 640 \\
	\hline
	Modulation & QPSK \\
	\hline
	Baseband sampling frequency (MHz) & 0.6 \\
	\hline
	Inter-carrier spacing (kHz) & 4.6875 \\
	\hline
\end{tabu}
\end{center}
\caption{System parameters.}
\label{tab:parameters}
\end{table}

Different relay selection policies are applied to the PLC system to acquire a quantitative comparison between the existing policies in the literature and the selection policies described in the proposed algorithms. In the first method, an exhaustive search is performed to find the best available relay node for transmission with the assumption of perfect CSI. This returns the highest reward at each time instance, or equivalently the maximum achievable reward for other algorithms which is a function of the channel conditions. In the second method, a fixed relay node is selected at the beginning of the transmission in a random manner and the selected node is used for the entire transmission time, regardless of the changes in the channel. In the third method the relay node is selected randomly for each instant of transmission. Therefore the all the available nodes have the chance of being selected, but the changes in channel does not play a role in the selection procedure. The fourth and the fifth methods of relay selection are the basic UCB and Discounted-UCB algorithms described in Section \ref{4-chapter}. Finally, the sixth and seventh selection policies are the proposed algorithms described in Section \ref{5-chapter}.

\begin{figure}[t]
\begin{center}
\epsfxsize=9cm \leavevmode\epsfbox{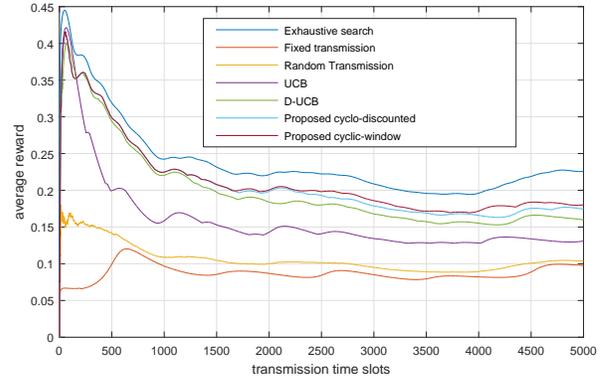} \caption{Average reward comparison between different relay selection policies.} \label{Fig:sim1}
\end{center}
\end{figure}

\begin{figure}[t]
\begin{center}
\epsfxsize=9cm \leavevmode\epsfbox{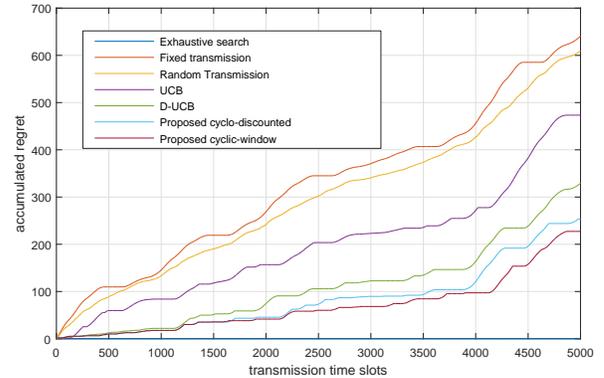} \caption{Accumulated regret comparison between different relay selection policies.} \label{Fig:sim2}
\end{center}
\end{figure}

Figure \ref{Fig:sim1} and \ref{Fig:sim2} show the average reward and the accumulated regret of the above mentioned relay selection policies, respectively. It can be seen that the best performance belongs to the exhaustive search method, as expected. The worst performance is from the random selection of the relay node and the fixed relay node, due to the fact that the changes in channel is disregarded in these methods. Basic UCB and discounted-UCB algorithms provide a significant performance gain in terms of average reward and demonstrate lower regrets throughout the transmission. It can be seen that the proposed algorithms can further improve the performance and obtain more reward and lower regret compared to the existing algorithms. Although, at the beginning of the operation, the padding functions are still large for all the arms and therefore a uniform weight as in UCB outperforms the discounted weights of the proposed algorithms.

\begin{figure}[t]
\begin{center}
\epsfxsize=9cm \leavevmode\epsfbox{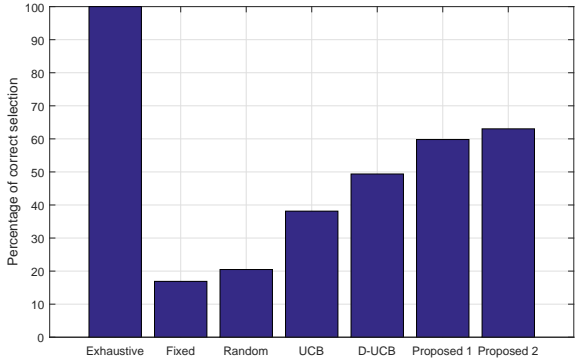} \caption{Percentage of correct selection of different transmission policies.} \label{Fig:sim3}
\end{center}
\end{figure}

Figure \ref{Fig:sim3} illustrates the percentage of correct selection in different transmission policies. This percentage is calculated after the entire duration of the transmission, by comparing the selected relay nodes of each method to the selected relay nodes of the exhaustive search method as a measure of reference. It can be seen that the basic MAB algorithms result in a higher percentage of correct selection and the proposed MAB algorithms can achieve yet higher percentages.

\begin{figure}[t]
\begin{center}
\epsfxsize=9cm \leavevmode\epsfbox{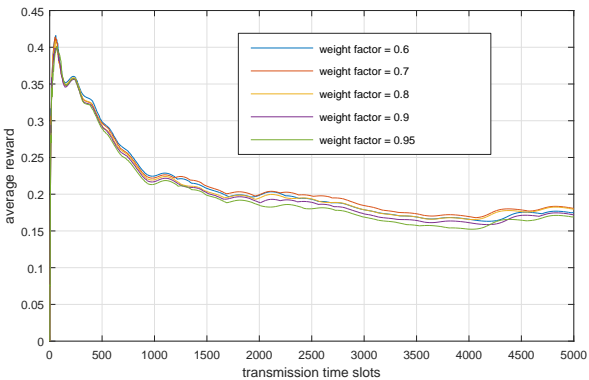} \caption{Impact of weight factor in the proposed cyclo-discounted UCB algorithm.} \label{Fig:sim4}
\end{center}
\end{figure}

\begin{figure}[t]
\begin{center}
\epsfxsize=9cm \leavevmode\epsfbox{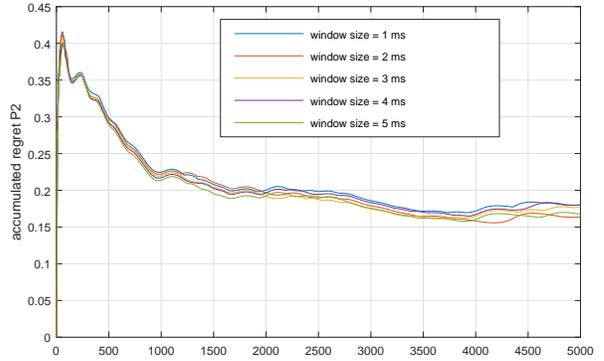} \caption{Impact of window size in the proposed cyclic-window UCB algorithm.} \label{Fig:sim5}
\end{center}
\end{figure}

\begin{figure}[t]
\begin{center}
\epsfxsize=9cm \leavevmode\epsfbox{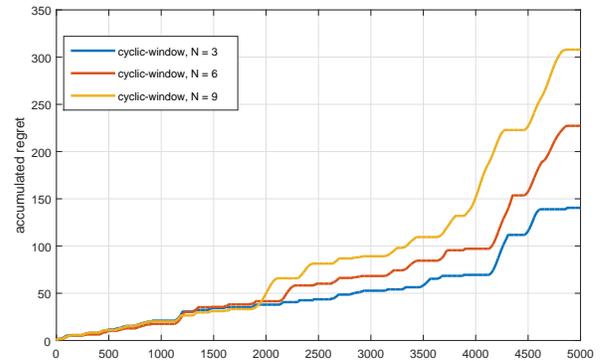} \caption{The proposed cyclic-window algorithm for different number of available relays.} \label{Fig:sim6}
\end{center}
\end{figure}

The first proposed algorithm, namely the cyclo-discounted UCB algorithm, uses a weight factor as a parameter of cyclic weights for the calculation of the UCB index, whereas the second proposed algorithm, namely the cyclic-window UCB algorithm, uses a window size parameter for the calculation of the UCB index. Figure \ref{Fig:sim4} and \ref{Fig:sim5} demonstrates the impact of different weight factors and window sizes, respectively. Compared with Figure \ref{Fig:sim1} we can realize that the change in window size or weight factor can hugely affect the performance of the algorithms. Figure \ref{Fig:sim6} demonstrates the effect of the number of relays on the cyclic-window algorithm. It can be seen that in the same frame of time, with higher the number of relays it takes longer for the algorithm to adapt itself to the environment and the accumulated regret grows by the number of available relays.

This paper has provided results of a two-hop relay selection scheme in PLC networks. However, the use of learning algorithms can be further generalized to multi-hop relay selection problems as well. In the case of $M$-hop transmission, with $N_i$, $i\in\{1,\cdots,M-1\}$ available relays at each hop, the number of available options raise to $\prod_{i=1}^{M-1} N_i$ which makes the time needed for finding the best path longer. Finding the best strategy for multi-hop relay selection can be regarded as a future work for this problem. Moreover, a generalized version of the proposed algorithm can be developed to further increase the reliability of the transmission. In this case, a multi-relay selection approach based on the learning algorithm is applied and at the receiver an appropriate combining technique is performed to further exploit the combining diversity among the best chosen relays. This strategy can further improve the reliability and decrease the probability of transmission in a less optimal route.

\vspace{10pt}
\section{\uppercase{Conclusion}}
\label{7-conclusion}

In this paper, we discussed the problem of relay selection in cooperative power line communication networks. The proper relay selection requires the availability of channel side information and the produced overhead of channel estimation makes it not realizable in a time-variant PLC environment. We introduced a class of machine learning algorithms, namely UCB algorithms in MAB model, to overcome this problem and improve the chances of selecting the best relay for transmission without channel side information. Furthermore, we proposed two new algorithms to further improve the selection policy by taking advantage of the periodicity of the synchronous impulsive noise of power line channels. We demonstrated through simulation results the supremacy of the proposed algorithms to the other policies of relay selection.

\section*{\uppercase{Acknowledgments}}
The authors thank S. Maghsudi for discussions on the use of upper confidence bound algorithm. The authors also thank anonymous reviewers for their constructive comments.

\bibliographystyle{jcn}

\epsfysize=3.2cm
\begin{biography}{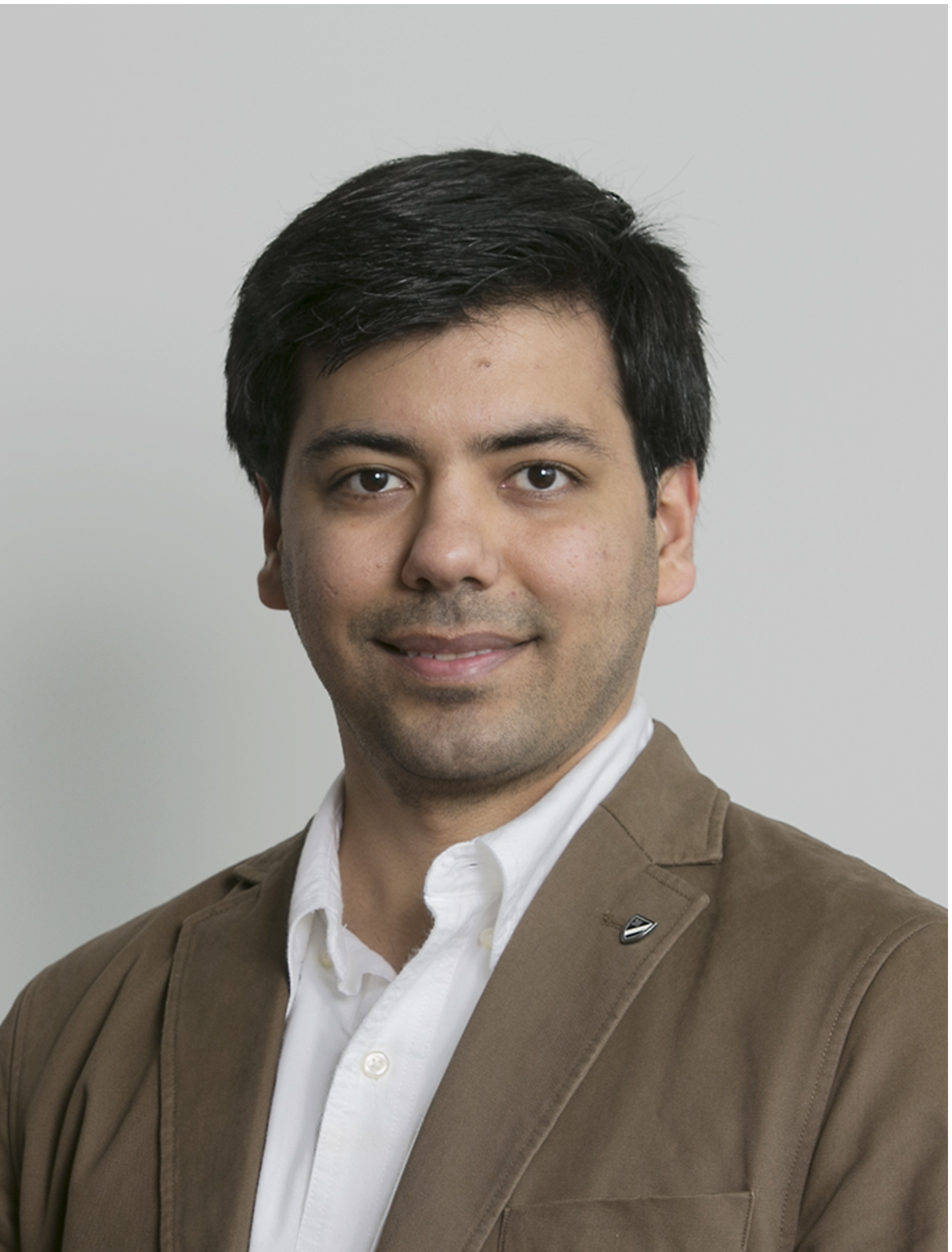}{Babak Nikfar} received his B.Sc. degree in Electrical Engineering with a major in Telecommunications from Iran University of Science and Technology, Tehran, Iran, in 2008 and the M.Sc. degree in Digital Communications from the University of Kiel, Kiel, Germany, in 2011. He is currently working toward the PhD degree with the Digital Communications Group, Department of Digital Signal Processing of the University of Duisburg-Essen, Duisburg, Germany.
\end{biography}

\epsfysize=3.2cm
\begin{biography}{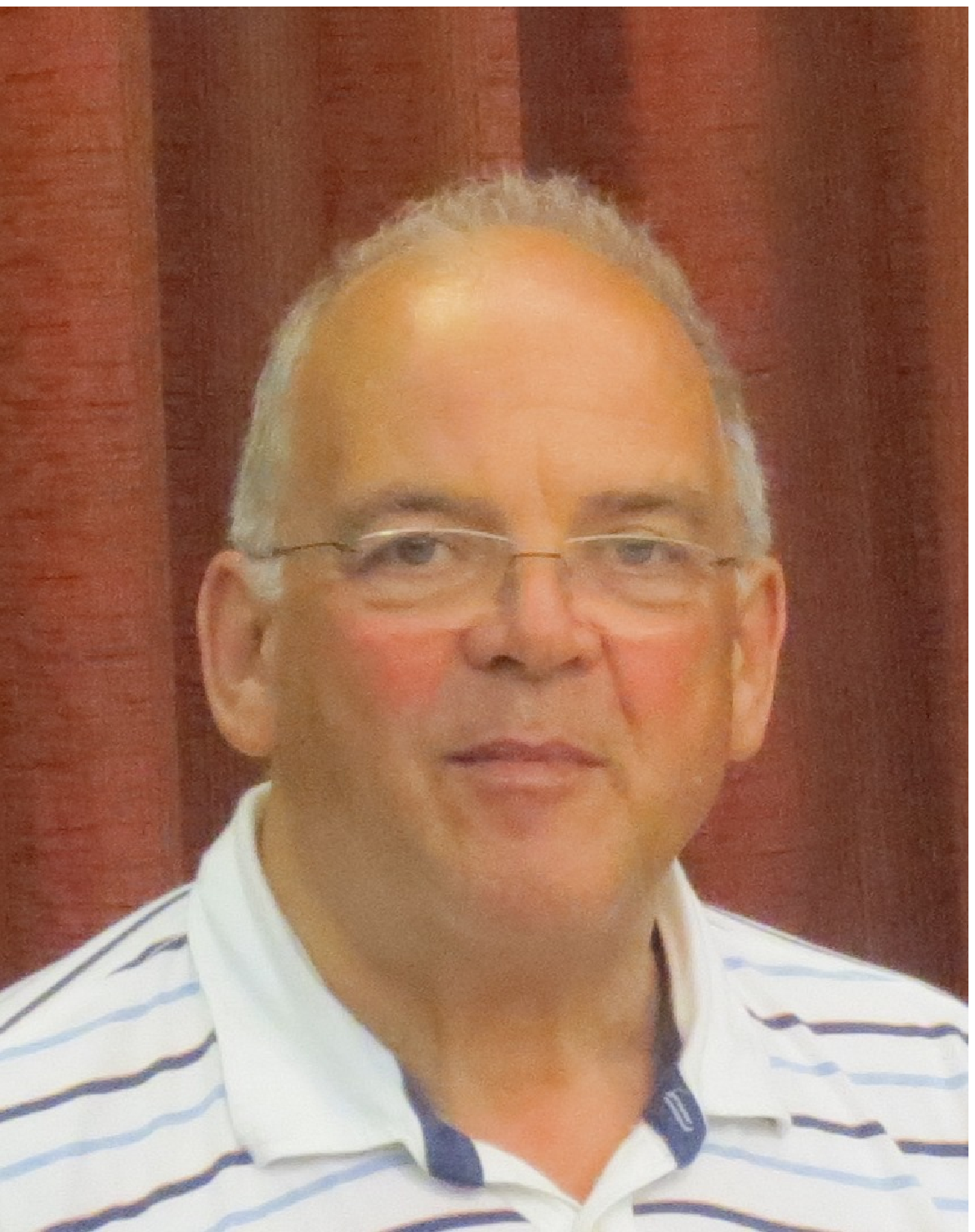}{A. J. Han Vinck}  is a senior professor in Digital Communications at the University
of Duisburg-Essen, Germany, specialising in Information and
Communication theory, Coding and Network aspects in digital communications.
  He held positions of professional responsibility including the Director of
the Institute for Experimental   Mathematics in Essen, founding Chairman of
the IEEE German Information Theory chapter, President of the IEEE
Information theory Society (2003) and President of the Shannon, Leibniz and
Gauss foundations for the stimulation of research in the field of
Information theory and Digital Communications. He received a number of
accolades including the election by the IEEE as Fellow for his
"Contributions to Coding Techniques", appointed as Distinguished Lecturer
for the Information Theory Society as well as for the Communications Society
of the IEEE, the IEEE ISPLC 2006 Achievement award for contributions to
Power Line Communications, the SAIEE annual award for the best paper
published in the SAIEE Africa Research Journal in 2008 and the 2015 Aaron D.
Wyner Distinguished Service Award for longstanding contributions to the IEEE
Information Theory society.
    He was instrumental in the organisation of research forums including the
IEEE Information Theory workshops and symposia Japan-Benelux workshops on
Information theory (now Asia-Europe workshop on "Concepts in Information
Theory") and the International Winterschool  on Coding, Cryptography and
Information theory in Europe.
    He is author of the book "Coding Concepts and Reed-Solomon Codes".
\end{biography}

\end{document}